\documentclass[11pt]{article}
\usepackage{moriond,epsfig}

\def\Journal#1#2#3#4{{#1} {\bf #2}, #3 (#4)}

\def\PRL{\em Phys. Rev. Lett.}
\def\PRD{{\em Phys. Rev.} D}

\def\be{\begin{equation}}
\def\ee{\end{equation}}
\def\bea{\begin{eqnarray}}
\def\eea{\end{eqnarray}}

\begin{document}
\vspace*{4cm}
\title{B PHYSICS AT THE TEVATRON RUN II}

\author{ K. YIP \footnote{On behalf of the CDF and D{\O} collaborations.} }

\address{Brookhaven National Laboratory, Bldg. 911B, Upton, NY 11973-5000, USA}

\maketitle\abstracts{
We present the B physics results from the CDF and D{\O} experiments
at the Tevatron Run II at Fermilab and their future prospect.  
This includes various B mass and lifetime measurements, B
mixing,  the confirmation of the discovery of
the X particle, rare decays, CP violation and spectroscopy.}

\section{Tevatron, the Experiments and Detectors}
\label{sec:prod}

After the shutdown in the autumn of 2003, the Tevatron has had significant
improvement in its luminosity delivery.  At the time of the conference,
the peak luminosity at Run II is above $7 \times {10}^{31} \rm cm^{-2} s^{-1}$
and both CDF and D{\O} experiments each has collected more than 290 $\rm {pb}^{-1}$
on tapes.  Both experiments have run quite stably and the data taking
efficiencies are about 80-90\%.   The results shown here include
analyses from the CDF data corresponding to about 65 to 220 $\rm {pb}^{-1}$ 
and the D{\O} data corresponding to about 115 to 250 $\rm {pb}^{-1}$.

Unlike B factories running at the $\Upsilon$(4S) resonance,
all hadronic states such as $\rm B_{s}$ and ${\Lambda}_{b}$
are produced at the Tevatron that
provides unique capabilities for studying these
particles and their interactions  in near future.  

%The ${b}\overline{b}$ production cross-section at the
%Tevatron is several orders of 
%magnitudes higher than the $e^{+}e^{-}$ machines, 
%but the inelastic scattering cross-section is even
%larger.  In order to 
%study b decays which belong to the lower end of the 
%energy and momentum spectra at Tevatron,
%specialized lepton and di-lepton triggers such as 
%$\rm J/\psi \rightarrow \mu^{+} \, \mu^{-}$
%are required.  In Run II, both the CDF and D{\O} experiments
%have moved the track triggers to Level 1.
%Precise secondary vertex reconstruction is also 
%necessary to have detailed study in B decays.

Detailed Run II detector upgrade information for CDF and D{\O} experiments
can be found in the reference~\cite{detector}.  
Both experiments at Run II have silicon vertex detectors,
axial solenoids providing magnetic fields for the central
tracking systems as well as competitive calorimeters and muon
detectors.

CDF has successfully implemented Level-2 trigger on displaced vertices
since the beginning of Run II.  CDF is stronger in particle 
identification (with the ``Time of Flight'' detector and dE/dx)
and has excellent mass resolution.
{D\O} excels in muon identification and tracking acceptance 
with the pseudo-rapidity coverage $|{\eta}| < 2-3$.  
{D\O} has the Level-3 trigger on impact parameter and the 
Level-2 impact parameter trigger is being commissioned.

\section{B Masses and Lifetimes}
\label{sec:mass_lifetime}

Both experiments have had very good yields of B hadrons.  
For example, {D\O} has so far reconstructed $\rm (4306 \pm 89)$
$\rm B^+ \rightarrow J/\psi +  K^+$, 
$\rm (1857 \pm 72)$ $\rm B^0 \rightarrow J/\psi + K^\star$
and
$\rm (375 \pm 29)$ $\rm B^0 \rightarrow J/\psi + K^0_s$ decays
with a data set corresponding to 225 $\rm {pb}^{-1}$.  After analyzing
about 115 $\rm {pb}^{-1}$ equivalence of data, {D\O} has also had a 
mass measurement of $\rm B^{\star \star}$ with 
$\rm (65 \pm 17)$ events in the peak and the mass difference
between $\rm B^{\star \star}$ and $\rm B^+$ is
$\rm (0.426 \pm 0.016) \ GeV/c^2$.

\begin{figure}[htp!]
\begin{center}
 \mbox{\epsfig{file=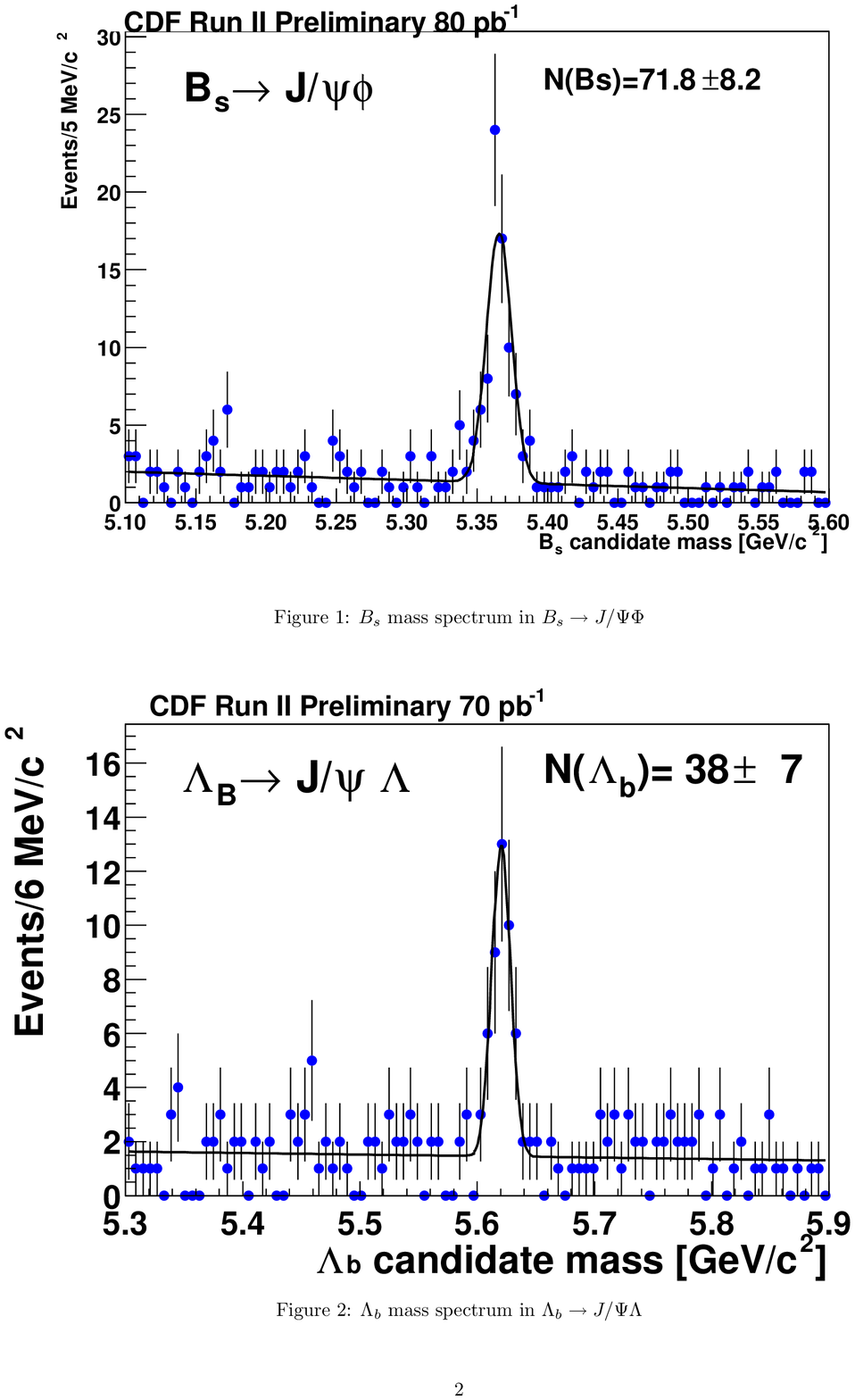,height=5.0cm,clip=, bbllx=81, bblly=471, bburx=494, bbury=757}} \hspace{0.88cm}
 \mbox{\epsfig{file=Bs_Lambda.ps,height=5.0cm,clip=, bbllx=41, bblly=92, bburx=501, bbury=408}}
\vspace{-0.28cm}
 \caption{The world's best mass measurements of $\rm B_s$ (left) and $\rm \Lambda_b$ (right) from the CDF experiment.}
 \label{fig:CDF_masses}
\end{center}
\end{figure}

CDF has already made the world's best measurements of $\rm B_s$ and $\rm \Lambda_b$ 
 (Fig.~\ref{fig:CDF_masses}):
\bea
\rm M(B_s) = \rm 5365.50 \pm 1.29 \ (stat) \ \pm 0.94 \ (syst) \ MeV/c^2 \ \ \ with \ 80 \ pb^{-1}\\
\rm M(\Lambda_b) = \rm 5620.4 \pm 1.6 \ (stat) \ \pm 1.2 \ (syst) \ MeV/c^2 \ \ \  with \ 70 \ pb^{-1}.
\eea
{D\O} has also made similar mass measurements for $\rm B_s$ and $\rm \Lambda_b$
and is going to reprocess with extended tracking to obtain higher
yield and perhaps better mass resolution.
{D\O} has nevertheless collected the world's largest sample of
exclusive $\rm B_s \rightarrow J/\psi + \phi$ decays with a yield of 
$\rm (403 \pm 28)$, corresponding to 225~$\rm {pb}^{-1}$.  

B lifetime measurement is a good meeting point between theories and
experiments as theoretical calculation of the widths of inclusive decays
remains a challenge.
Both experiments have engaged in lifetime measurements of B hadrons.
The most recent results are tabulated in Table~\ref{tab:lifetime}.
\begin{table}[h]
\caption{$\rm B_s$ Lifetime measurements ($\rm \tau_{B_s}$) and their 
ratios to the $\rm B^0$ lifetimes ($\rm \tau_{B^0}$)
from CDF and {D\O}.\label{tab:lifetime}}
\begin{center}
\begin{tabular}{|c|c|c|}
\hline
&  {CDF} &  {D\O}  \\ \hline
Corresponding luminosities &  $\rm 138 \ {pb}^{-1}$ &  $\rm 115 \ {pb}^{-1}$ \\ \hline
$\rm \tau_{B_s}$  &
$\rm 1.330^{\ + 0.148}_{\ - 0.129} \ (stat) \pm 0.02 \ (syst) \ ps$ & 
$\rm 1.190^{\ + 0.19}_{\ - 0.16} \ (stat) \pm 0.14 \ (syst) \ ps$   \\ \hline
$\rm \tau_{B_s}/\tau_{B^0}$ & $\rm  0.89 \pm 0.10 $ &  $\rm  0.79 \pm 0.14$ \\ \hline
\end{tabular}
\end{center}
\end{table}

\begin{figure}[htp!]
\begin{center}
% \mbox{\epsfig{file=B03F04_pretty.eps,width=7.601cm,clip=}}
 \mbox{\epsfig{file=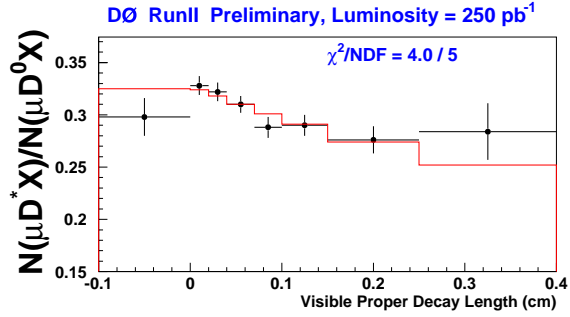,width=7.5cm,clip=}}
\vspace{-0.28cm}
 \caption{The ratio of events in $\rm D^{\star}$ and 
$\rm D^0$ samples as a function of the visible proper decay length.  
The dots are the data points and the line is the fit from the minimization of $\rm \chi^2$.}
 \label{fig:BplusB0ratio}
\end{center}
\end{figure}

It is most convenient to study lifetime ratios because in this way one cancels
the dependence on quantities which are poorly known, such as the b-quark mass.
This is not only true in theoretical calculations but also in experimental
measurements.  Instead of measuring lifetimes separately and taking their ratios,
{D\O} has recently measured the lifetime ratio of $\rm B^+$ and $\rm B^0$ 
directly from the fit of the ratio of the number of events between
$\rm B \rightarrow \mu^+ \nu D^{\star} X$ and
$\rm B \rightarrow \mu^+ \nu D^0 X$ decays (~where X implies other particles~)
as a function of visible proper decay lengths (Fig.~\ref{fig:BplusB0ratio}).
In this analysis, sample compositions, measured branching fractions and
isospin relations are used.  Full detector simulations 
are used to extract the K-factors, relative reconstruction efficiencies
for different B decay modes and decay length resolutions. The lifetime of
$\rm B^+$ has been fixed to be $\rm (1.674 \pm 0.018)$~ps \, \cite{PDG}.
The preliminary result for this lifetime ratio with a data set
from $\rm 250 \ {pb}^{-1}$ is 
\be 
\rm \frac{\tau(B^+)}{\tau(B^0)} = 1.093 \pm 0.021 \ (stat) \pm 0.022 \ (syst)
\ee
This is one of the most precise B lifetime ratio measurements in the world.

With a data set from $\rm 65 \ {pb}^{-1}$,
CDF has also measured the lifetime of $\rm \Lambda_b$:
\be 
\rm c \tau = 374 \pm 78 \ (stat) \pm 29  \ (syst) \ \mu m \ .
\ee
The largest systematic error is from asymmetric track reconstruction
in the Central Outer Tracker which is about 26 $\rm \mu$m.  Work is
in progress to add more data and use better track reconstruction.
The lifetime measurement of $\rm \Lambda_b$ in {D\O} is also in 
progress.

\section{B Mixing}
\label{sec:mixing}

{D\O}, in addition to CDF, has recently demonstrated its capability to measure B mixing.
A large sample of semi-leptonic $ B^0 / \overline{B^0} $ decays, corresponding
to $\rm 250 \ {pb}^{-1}$, has been used to measure the mass difference
between $B_d$ and $\overline{B_d}$, $\Delta m_d$, 
which is also called the oscillation frequency:
\be
\rm \Delta m_d = 0.506 \pm 0.055 \ (stat) \pm 0.049  \ (syst) \ {ps}^{-1} \ .
\ee
It is consistent with the world average of 
$\rm (0.502 \pm 0.007) \ {ps}^{-1}$.  This analysis uses opposite-side
muon tag.  The tagging efficiency is $\rm (4.8 \pm 0.2) \%$ and the
tagging purity is $\rm (73.0 \pm 2.1) \%$.  Work is now in progress
in {D\O} to use other tagging methods and include more decay channels.

The oscillation frequency for $\rm B_s$, $\rm \Delta m_s$,  is expected to be
much larger than $\Delta m_d$.    This
 makes  $\rm \Delta m_s$ more difficult to measure than $\rm \Delta m_d$. 
Nevertheless, the Tevatron is the only place which can measure  $\rm \Delta m_s$
in near future.  
There are generally two approaches to measure $\rm \Delta m_s$.  

The first
approach is through the semi-leptonic decays.  This approach has
very good statistics but the time resolution is poor due to the undetected
neutrinos in the decays.  If  $\rm \Delta m_s$ is about $\rm 15 \ {ps}^{-1}$,
it is expected that CDF and {D\O} each can have a 1 to 2 $\rm \sigma$ measurement
with a luminosity of $\rm 500 \ {pb}^{-1}$.

The other approach makes use of the fully reconstructed hadronic decays.
Though the statistics here are poorer, the time resolution is excellent.
It is expected that we need a few $\rm fb^{-1}$ of data to measure 
$\rm \Delta m_s$ to about 5 $\rm \sigma$ if $\rm \Delta m_s$
is about  $\rm 18 \ {ps}^{-1}$.

The ``golden channel'' for CDF is $\rm B_s \rightarrow D_s \, \pi$. Here
the proper time resolution at CDF is about 67 fs and the B momentum resolution
is about 0.5\%.  CDF expects to have a reconstruction yield about $\rm 0.7/{pb}^{-1}$ 
and the signal to background ratio is about 2.  Combining a few analyses,
CDF has measured the following ratio:
\be
\rm \frac{f_s}{f_d} \cdot \frac{Br(B_s \rightarrow D_s^- {\pi}^+)}{Br(B^0 \rightarrow D^- {\pi}^+)} =
0.35 \pm 0.05 \ (stat) \pm 0.04 \ (syst) \pm 0.09 \ (Br)
\ee
$\rm f_d$ and $\rm f_s$ are the fractions of $\rm B_d$ and $\rm B_s$ in the data 
sample respectively.

\section{X(3872), Rare Decays, CP violation and Spectroscopy}
\label{sec:miscellaneous}

Both CDF and {D\O} have confirmed the discovery of the X(3872) by 
the Belle Collaboration~\cite{X} and measured compatible mass values.  
This shows that X particles are not only from B decays but also produced
promptly.  The large detector acceptance allows {D\O} to show that
the X particles are produced both in the central ($\rm |\eta| < 1$)
as well as in the forward regions ($\rm 1 < |\eta| < 2$).  
From the isolation and decay length comparisons with $\rm \psi(2S)$, 
the production of the X particles has the same mixture of prompt
and long-lived fractions as the $\rm \psi(2S)$.  

The decay of $\rm B_s \rightarrow \mu^+ \mu^-$ is highly suppressed 
(Br($\rm B_s \rightarrow \mu^+ \mu^-$) $\rm \sim {10}^{-9}$)
in the Standard Model whereas it has about 2 orders of magnitude 
enhancement in the Supersymmetry Model.  CDF has performed a
blind analysis and obtained the following limits at 90\% confidence level:
\bea
\rm Br(B_s \rightarrow \mu^+ \mu^-) < 5.8 \times {10}^{-7} \ ,\\
\rm Br(B_d \rightarrow \mu^+ \mu^-) < 1.5 \times {10}^{-7} \ .
\eea
These are the best limits published for both decay channels.
{D\O} is also carrying out a blind analysis. Since
{D\O} tries to optimize the analysis without bias, the
signal region is still hidden.  A current sensitivity study at
{D\O} shows that a limit of 
$\rm Br(B_s \rightarrow \mu^+ \mu^-) < 1 \times {10}^{-6}$ at
95\% confidence level can be reached and further improvement 
is expected.

Another main topic is the studies of CP violation by measuring
various angles in the unitary triangle related to the CKM matrix
as well as measuring
the direct CP asymmetry from the self-tagging modes such as
$\rm B_d \rightarrow K \, \pi $ and $\rm B_s \rightarrow K \, \pi$.
Displaced track trigger at Level-2 gives CDF accessibility to
rare hadronic decays with high signal to background ratio.
dE/dx in CDF provides a statistical separation of kaons from
protons at the level of 1.3~$\sigma$.

Finally, it is also possible to study spectroscopy at the Tevatron.
In the sample of the decay channels
 $\rm B \rightarrow \mu \, \nu \, D^{\star +} \, X$, {D\O} tries to
reconstruct another $\rm \pi^-$ and look at the invariant mass
distribution of the $\rm D^{\star +} \, \pi^-$ system. A clear
excess is seen and can be interpreted as an interfering 
Breit-Wigner function of the narrow states $\rm D_1^0$
and $\rm D^{\star 0}_2$ which decays through D-waves.
Assuming
the branching fraction
$\rm Br(B \rightarrow \, \mu \, \nu \, D^{\star +} \, \pi^- \, X) = (0.48 \pm 0.10) \%$
from the available LEP results from topological analyses~\cite{PDG}, 
the preliminary result of {D\O} for the
first time constrains the resonant contribution
\be
\rm Br(B \rightarrow \{D^0_1,D^{\star 0}_2\} \mu \, \nu \, X) \cdot 
    Br(\{D^0_1,D^{\star 0}_2\}  \rightarrow   D^{\star +} \, \pi^-)  
= 0.280 \pm 0.021 \, (stat) \pm 0.088 \, (syst) \, \% \, .
\ee
Work is in progress to extract separate amplitude and phase for
each state.

\section*{Acknowledgments}
The author acknowledges the useful suggestions from the CDF and D{\O} colleagues
such as C. Paus, V. Jain, R. van Kooten, A. Zieminski and D. Zieminska during the 
preparation for the presentation in this conference.

\end{document}